\documentclass[%
aip,
jmp,%
amsmath,amssymb,
reprint,%
]{revtex4-2}

\usepackage{amsmath,graphicx,array}
\usepackage{dcolumn,soul}%

\usepackage{amsthm}
\usepackage[figuresright]{rotating}%
\usepackage{algorithm, algorithmicx, algpseudocode}
\usepackage{listings}%
\usepackage{hyperref}
\usepackage{bm}

\makeatletter
\def\uns{\ifmmode\,\else$\,$\fi}%

\makeatother
\usepackage[fontsize=12pt]{fontsize}
 
\begin{document}
	
	\preprint{AIP/123-QED}
	
	\title{Ultralow shot noise limited giant passive resonant gyroscope for Earth rotation measurement}
	
	\author{Yuhong Zhong}%
	\author{Yangsheng Cai} 
	\author{Zhanhao Liu}%
	\author{Lei Zheng}%
	\author{Yunhe Wang}%
	\author{Xiaojun Huang}%
	\author{Zhiyuan Wang}%
	\affiliation{%
		National Gravitation Laboratory, MOE Key Laboratory of Fundamental Physical Quantities Measurement and School of Physics, Huazhong University of Science and Technology,\\  Wuhan 430074, People's Republic of China
	}%
	\author{Kui Liu}%
	\author{Liangcheng Tu}%
	\author{Jun Luo}
	\affiliation{%
		MOE key Laboratory of TianQin Mission, TianQin Research Center for Gravitational Physics \& School of Physics and Astronomy, Frontiers Science Center for TianQin, Gravitational Wave Research Center of CNSA, Sun Yat-sen University (Zhuhai Campus),\\ Zhuhai 519082, People’s Republic of China
	}%
	\author{Zehuang Lu}%
	\author{Jie Zhang}\altaffiliation[jie.zhang@mail.hust.edu.cn]{}%
	\affiliation{%
		National Gravitation Laboratory, MOE Key Laboratory of Fundamental Physical Quantities Measurement and School of Physics, Huazhong University of Science and Technology,\\  Wuhan 430074, People's Republic of China
	}%
	
	\date{\today}
	
	\begin{abstract}
		Optical gyroscopes directly measure the Earth's rotation and are promising instruments for real-time geophysical observations and Earth orientation parameter (EOP) determination requiring both high precision and high temporal resolution. Large-scale ring laser gyroscopes (RLGs) currently reach rotational resolutions around $10^{-11}\,\mathrm{(rad/s)/\sqrt{Hz}}$, but their quantum noise limits make it challenging to meet the requirements of future high-temporal-resolution EOP measurements. Passive resonant gyroscopes (PRGs), on the other hand, offer a potentially lower photon shot noise limit and more flexible power scaling, even if their demonstrated rotational resolutions are still about two orders of magnitude below those of leading RLGs. Here we demonstrate a $64\,\mathrm{m^{2}}$ giant passive resonant gyroscope HUST-2, and develop with an extremely low shot noise level. We experimentally obtain a shot noise limited of $5.7(1)\times10^{-13}\,\mathrm{(rad/s)/\sqrt{Hz}}$ at $1\,\mathrm{mW}$ incident optical power, following the characteristic $1/\sqrt{P}$ scaling. Through systematic suppression of dominant technical noise sources, HUST-2 further achieves a measured rotational resolution of $3\times10^{-11}\,\mathrm{(rad/s)/\sqrt{Hz}}$, bringing PRGs into the performance regime of leading large-scale RLGs for the first time. The gap between the present demonstrated rotational resolution and the shot noise limit indicates nearly two orders of magnitude further improvement potential. Reaching this limit would enable high-precision length-of-day (LOD) measurements with $10$-$100\,\mathrm{s}$ temporal resolution and lays the foundation for future large-scale gyroscope networks dedicated to real-time EOP determination.
	\end{abstract}
	
	\maketitle

\section{Introduction}\label{sec1}
Precise knowledge of the Earth's rotation is essential for modern space and time infrastructure~\cite{Wahr1988EarthRotation,altamimi_itrf2020_2023,charlot_third_2020,Petit_2010}. 
Earth orientation parameters (EOPs), including UT1-UTC, polar motion, and celestial pole offsets, define the transformation between the terrestrial and celestial reference frames and are indispensable for satellite navigation, deep-space tracking, precise orbit determination, geodesy, and fundamental astronomy~\cite{bizouard_iers_2019}. The most accurate EOP products are generated from combined analyses of space geodetic techniques, including very long baseline interferometry (VLBI), satellite laser ranging (SLR), global navigation satellite systems (GNSS), and Doppler Orbitography and Radiopositioning Integrated by Satellite (DORIS)~\cite{gambis_monitoring_2004,schuh_vlbi_2012}. VLBI is particularly important for UT1 and celestial pole offsets because it directly links the terrestrial station network to the extragalactic celestial reference frame~\cite{nothnagel_international_2017,schartner_optimal_2021}. However, final combined EOP products are not available in real time because global observations, data transfer and combined analysis require time. Rapid products reduce this delay by using prediction, but this generally degrades their accuracy relative to final products~\cite{kalarus_achievements_2010}. 
Therefore, real-time Earth rotation information with both high accuracy and high-temporal-resolution remains difficult to obtain.

Large optical gyroscopes provide a complementary route to Earth rotation metrology~\cite{schreiber_invited_2013}. 
A ground-based optical gyroscope directly senses the projection of the Earth's rotation vector onto its sensitive axis, enabling local, continuous, and real-time measurements without intrinsic dependence on external astronomical references~\cite{schreiber_variations_2023,song_advanced_2023,chen_giant_2025}. Large-scale active ring laser gyroscopes (RLGs) such as the G-ring have supported studies of Earth rotation variations, auxiliary EOP determination and investigations of signals associated with precession and nutation~\cite{nilsson_combining_2012,bohm_earth_2019,schreiber_gyroscope_2025}. 
The four-axis ROMY RLG has further demonstrated reconstruction and real-time output of the Earth rotation vector~\cite{gebauer_reconstruction_2020}. 
These results establish large optical gyroscopes as a viable complementary tool for high-precision Earth rotation monitoring. However, achieving subcentimeter level global geodetic point positioning in the terrestrial reference frame requires LOD estimates with an accuracy well below $0.1\,\mathrm{ms}$, which corresponds to resolving rotation rate fluctuations of $\Delta \Omega \simeq 8.4\times10^{-14}\,\mathrm{rad/s}$, i.e., on the order of $10^{-13}\,\mathrm{rad/s}$~\cite{gambis_earth_2011}. The state-of-the-art demonstrated rotational resolutions of large-scale RLGs, expressed as root power spectral densities (PSDs) of the gyroscope output in equivalent rotation rate, remain in the $10^{-11}\,\mathrm{(rad/s)/\sqrt{Hz}}$ regime~\cite{igel_romy_2021,di_virgilio_noise_2024,schreiber_gyroscope_2025}. Resolving $0.1\,\mathrm{ms}$-level LOD fluctuations therefore typically requires integration times of $10^4$-$10^5\,\mathrm{s}$. This limits the temporal resolution of EOP measurements and places stringent demands on long-term instrumental stability. Further improvement is increasingly constrained by the active ring cavity architecture of RLGs, which limits further reduction of the quantum noise floor through higher intracavity power or continued scaling of the cavity size~\cite{liu_active_2026}.

Passive resonant gyroscopes (PRGs) provide a fundamentally different route toward this performance frontier. 
By decoupling the passive sensing cavity from optical gain medium, PRGs allow more flexible laser power scaling and can, in principle, achieve a lower shot noise limit than that of RLGs~\cite{ezekiel_passive_1977}. 
In practice, however, this potential has not yet been realized. 
Existing PRGs have been limited by technical noise sources, including environmental perturbations, cavity length drift and laser frequency noise. 
As a result, typical PRGs currently exhibit rotational resolutions in the $10^{-7}$ to $10^{-9}\,\mathrm{(rad/s)/\sqrt{Hz}}$ range~\cite{martynov_passive_2019,korth_passive_2016,liu_large-scale_2019,feng_three-wave_2023,zenner_hansch-couillaud_2026}, leaving a performance gap of more than two orders of magnitude relative to that of the state-of-the-art RLGs.

In this work, we bridge this performance gap by demonstrating a rotational resolution of $3 \times 10^{-11}\,\mathrm{(rad/s)/\sqrt{Hz}}$ with HUST-2, a $64\,\mathrm{m^2}$ giant passive resonant gyroscope. 
This performance is achieved through systematic optimization of environmental effect isolation, scale factor stability and the frequency locking system. HUST-2 is the first PRG to reach the rotational resolution of the state-of-the-art large-scale RLGs. 
More importantly, by developing a quantitative shot noise model and experimentally characterizing the locking system, we demonstrate a shot noise limited of $5.7(1) \times 10^{-13}\,\mathrm{(rad/s)/\sqrt{Hz}}$ at $1\,\mathrm{mW}$ incident optical power. This noise limit corresponds to the few-$10^{-13}\,\mathrm{(rad/s)/\sqrt{Hz}}$ regime relevant to high time resolution EOP monitoring discussed above, and therefore provides a concrete route toward optical-gyroscope-based LOD measurements on $10$-$100\,\mathrm{s}$ timescales. 
The agreement between experiment and model further indicates that nearly two orders of magnitude improvement beyond the present demonstrated performance is achievable by suppressing technical noise. Such a low noise floor could provide high-temporal-resolution monitoring of Earth rotation variations, support future optical gyroscope EOP determination, and contribute to terrestrial tests of relativistic rotational effects\cite{schreiber_how_2011,di_virgilio_underground_2020}.

\section{Results}\label{sec2}
\subsection{Ultralow shot noise limit performance at the $10^{-13}\,\mathrm{(rad/s)/\sqrt{Hz}}$ level}

\begin{figure*}[htb]
	\centering
	\includegraphics[width=0.95\textwidth]{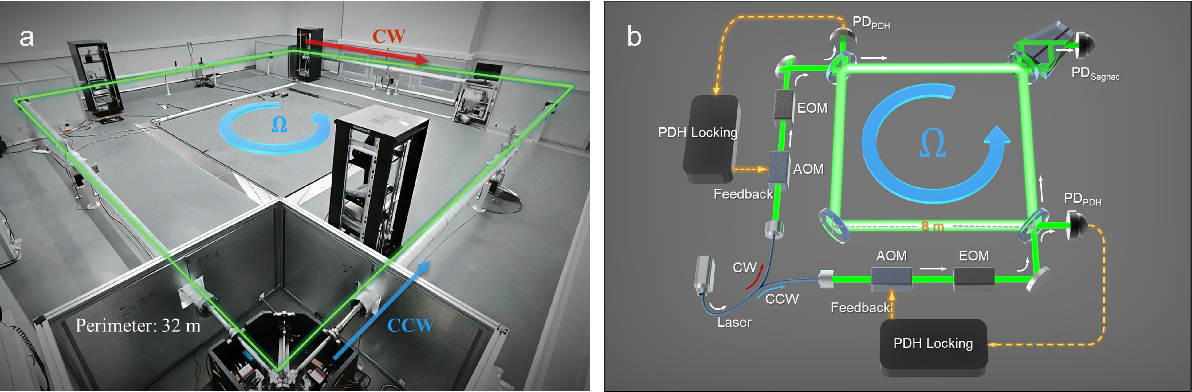}
	\caption{\label{fig:sys} \textbf{The HUST-2 PRG.} \textbf{a,} Photograph of HUST-2. \textbf{b,} Schematic diagram of the HUST-2 PRG. The external $532\,\mathrm{nm}$ laser is symmetrically injected into the $8\,\mathrm{m} \times 8\,\mathrm{m}$ ring cavity. Bidirectional PDH loops, utilizing EOMs and AOMs, tightly track the ultra narrow resonances. The rotation signal $\Omega$ is extracted from the beat note at the $\mathrm{PD}_{\mathrm{Sagnac}}$.}
\end{figure*}

In a resonant optical gyroscope, the rotation induced Sagnac frequency splitting $f_{\rm S}$ between the clockwise (CW) and counterclockwise (CCW) modes is given by:
\begin{equation}
	f_{\rm S} = \bm{K} \cdot \bm{\Omega}_{\rm E} = \frac{4A}{\lambda L}\Omega_{\rm E} \cos\theta,
\end{equation}
where $\bm{K}, \bm{\Omega}_{\rm E}$ are the scale factor and Earth rotation rate vectors, respectively. $A$ is the enclosed area, $L$ is the perimeter, $\lambda$ is the operating wavelength, and $\theta$ is the colatitude of gyroscope location~\cite{schreiber_invited_2013}. Maximizing the magnitude of $K$ is essential for improving sensor resolution. Unlike RLGs, PRGs physically isolate the laser source from the ring cavity. This architecture avoids complex intracavity gain dynamics, enabling larger cavity dimensions, and the use of shorter optical wavelengths. Leveraging this advantage, we construct HUST-2: a large-scale PRG featuring an $8\,\mathrm{m} \times 8\,\mathrm{m}$ square cavity interrogated by a $532\,\mathrm{nm}$ laser, as shown in Fig.~\ref{fig:sys}a. By combining a $32\,\mathrm{m}$ perimeter with ultralow loss mirrors, the optical loss is minimized, yielding an optical $Q$-factor of $6.4 \times 10^{12}$, the best reported value so far. More additional key parameters are given in the Methods section.

As depicted in Fig.~\ref{fig:sys}b, HUST-2 operates on independent, bidirectional Pound-Drever-Hall (PDH) locking of the cavity resonances. To suppress frequency fluctuations of the laser, the external $532\,\mathrm{nm}$ laser is pre-locked to the ring cavity~\cite{feng_three-wave_2023}. The beam is then split, phase modulated via electro-optic modulators (EOMs), and symmetrically injected into the CW and CCW paths. Independent servo loops process the reflected PDH error signals and drive acousto-optic modulators (AOMs) to tightly lock the incident laser frequencies to the same cavity mode. The Sagnac signal is extracted from the beat note of the transmitted fields at the $\mathrm{PD}_{\mathrm{Sagnac}}$. Given the scale factor of HUST-2 and the local latitude ($30^\circ 31' \, \mathrm{N}$), the Earth induced Sagnac frequency $f_{\rm S}$ is approximately $557\,\mathrm{Hz}$.

Because the PRG is driven by external laser injection, its fundamental limit is not governed by gain medium dynamics, as in active RLG. Instead, the ultimate performance limit is set by the photon shot noise associated with optical detection. As shown in Fig.~\ref{fig:sys}b, the experimental system contains three primary PDs. Among them, the shot noise introduced in the PDH frequency discrimination process gives the dominant contribution to the fundamental rotation noise. The detailed derivation is given in the Methods.

To quantify this limit, we model how the detection shot noise propagates through the frequency stabilization loop. For one propagation direction, the shot noise induced voltage fluctuation in error signal is

\begin{equation}
	\sqrt{S_V^{\mathrm{sn}}}
	=G_{\mathrm{el}}\sqrt{2eI_{\mathrm{PD}}}
	=G_{\mathrm{el}}\sqrt{\frac{2\eta e^2}{h\nu} R P_{\mathrm{in}}},
	\label{eq:shot_noise_voltage}
\end{equation}
where $G_{\mathrm{el}}$ is the gain of electronic chain, $e$ is the elementary charge, $I_{\mathrm{PD}}$ is the photocurrent, $\eta$ is the detector quantum efficiency, $h\nu$ is the photon energy, $P_{\mathrm{in}}$ is the incident optical power, and $R=P_{\mathrm{refl}}/P_{\mathrm{in}}$ is the on-resonance reflection ratio. Here $P_{\mathrm{refl}}$ denotes the reflected optical power incident on the $\mathrm{PD_{PDH}}$.

The corresponding frequency fluctuation is obtained by normalizing the voltage noise to the linear PDH frequency discrimination slope $D$, given by $\sqrt{S_\nu^{\mathrm{sn}}}= \sqrt{S_V^{\mathrm{sn}}}/D$. For the PDH error signal used here, the frequency discrimination slope can be written as~\cite{black_introduction_2001}

\begin{equation}
	D =
	G_{\mathrm{el}}
	\frac{8J_0J_1\alpha(1-\zeta)\eta e P_{\mathrm{in}}}
	{\nu_{\rm c}h\nu},
	\label{eq:pdh_slope}
\end{equation}
where $J_0$ and $J_1$ are the Bessel functions determined by the phase modulation index, $\alpha,\zeta$ represent the spatial mode-matching efficiency and cavity impedance-matching coefficient, respectively. $\nu_{\rm c}$ is the cavity linewidth.

In a PRG, the rotation signal is extracted from the frequency difference between the CW and CCW beams, $f_{\mathrm{S}} = \nu_{\mathrm{cw}}-\nu_{\mathrm{ccw}}$. Since the two beams are independently locked to the cavity, their shot noise contributions are statistically uncorrelated. Under symmetric operating conditions, the shot noise PSD of the frequency difference is given as $S_{f_{\mathrm{S}}}^{\mathrm{sn}}=S_{\nu,\mathrm{cw}}^{\mathrm{sn}}+S_{\nu,\mathrm{ccw}}^{\mathrm{sn}}\approx2S_{\nu}^{\mathrm{sn}}$. Thus, the frequency noise root PSD of the Sagnac frequency signal is increased by a factor of $\sqrt{2}$. Dividing this noise by the scale factor $K=4A/(\lambda L)$, we obtain the rotation equivalent shot noise limit

\begin{equation}
	\sqrt{S_\Omega^{\mathrm{sn}}}
	=
	\frac{\lambda L}{4 A}
	\cdot
	\frac{\nu_{\rm c}}{4 J_0 J_1 \alpha (1-\zeta)}
	\sqrt{\frac{h\nu R}{\eta P_{\mathrm{in}}}} .
	\label{eq:sn_realistic}
\end{equation}

Equation~(\ref{eq:sn_realistic}) gives the shot noise limited of the PRG. It shows that the rotation noise scales inversely with the square root of the incident optical power, $\sqrt{S_\Omega^{\mathrm{sn}}}\propto 1/\sqrt{P_{\mathrm{in}}}$. This scaling provides a direct route to improving the noise floor by increasing the injected optical power. In addition, reducing the cavity linewidth $\nu_{\rm c}$, or equivalently increasing the cavity quality factor, can further lower the shot noise limit.

\begin{figure}[htb]
	\centering
	\includegraphics[width=0.6\linewidth]{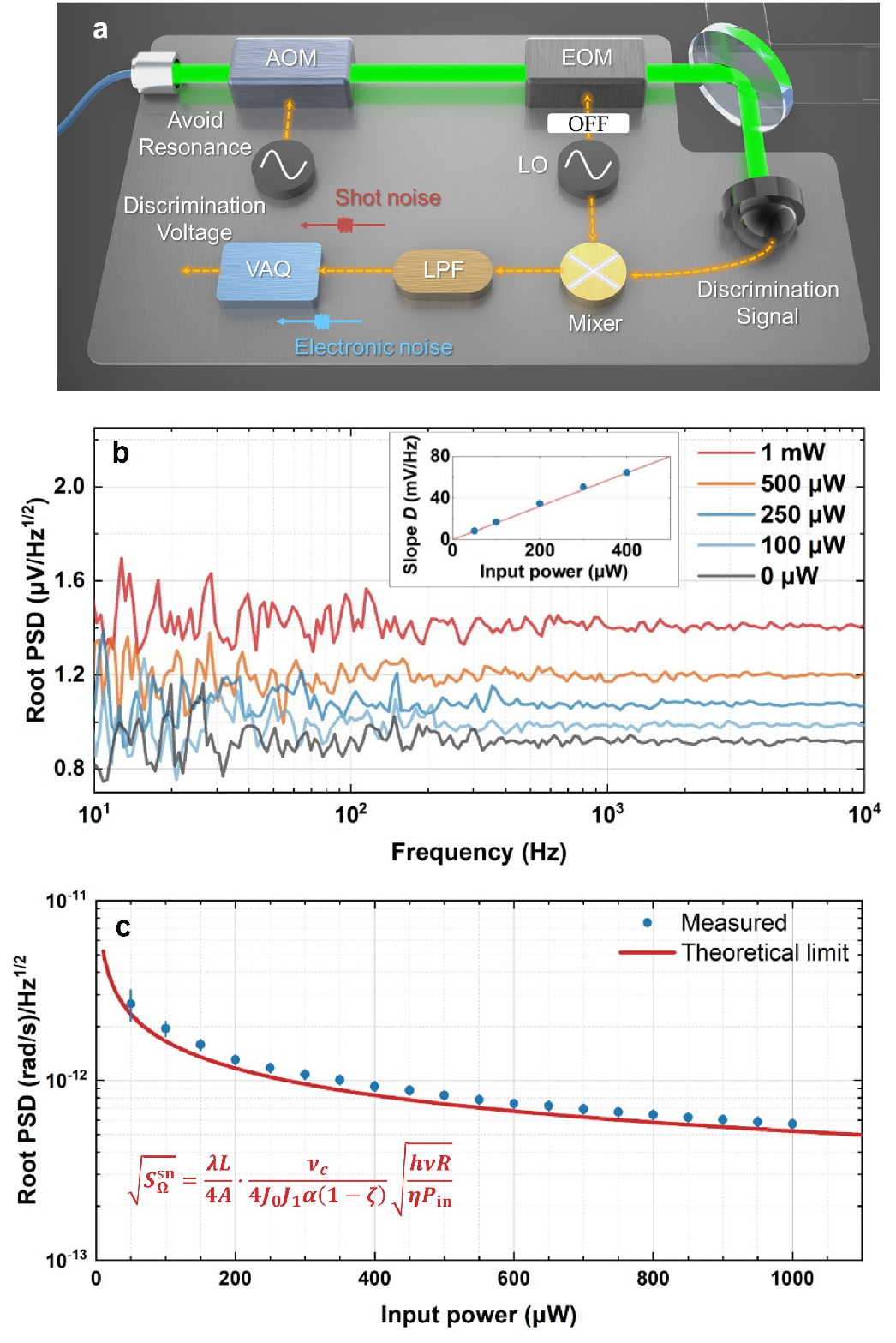}
	\caption{\textbf{Experimental extraction of the fundamental shot noise limit and future roadmap.} 
		\textbf{a,} Schematic of the variable control experiment designed to isolate pure optical shot noise. 
		\textbf{b,} Extracted voltage noise root PSDs at various incident optical powers, exhibiting flat white-noise characteristics. The inset shows the linear dependence of the measured discrimination slope $D$ on the input power. 
		\textbf{c,} The measured shot noise limited (blue dots) as a function of input power, demonstrating good agreement with the theoretical quantum limit (red solid line). At $1\,\mathrm{mW}$ input power, the fundamental limit reaches $5.7 \times 10^{-13}\,\mathrm{(rad/s)/\sqrt{Hz}}$. }
	\label{fig:shot_noise_results}
\end{figure}

This explicit power dependence also provides a direct experimental signature of the shot noise model. To verify it, we vary the incident optical power and measure the corresponding rotation equivalent noise, which is then compared with the $1/\sqrt{P_{\mathrm{in}}}$ scaling predicted by Eq.~(\ref{eq:sn_realistic}). However, the shot noise limit cannot be directly observed in the locked system, because the PDH frequency discrimination chain also contains several technical noise contributions. These noises couple into the error signal and appear as equivalent frequency, and hence rotation, fluctuations. We therefore express the frequency discrimination equivalent rotational noise in one loop as
\begin{equation}
	\sqrt{S_\Omega^{\mathrm{freq}}}
	=
	\frac{1}{DK}
	\sqrt{
		S_V^{\mathrm{sn}}
		+
		S_V^{\mathrm{el}}
		+
		S_V^{\mathrm{RAM}}
		+
		S_V^{\mathrm{bs}}
	},
	\label{eq:somega_dis}
\end{equation}
where $S_V^{\mathrm{el}}$, $S_V^{\mathrm{RAM}}$, and $S_V^{\mathrm{bs}}$ denote the electronic noise, residual amplitude modulation (RAM) noise, and backscattering noise, respectively.

To overcome this, we implement a systematic measurement procedure, as illustrated in Fig.~\ref{fig:shot_noise_results}a, to step-by-step isolate and subtract out these technical noise sources. First, an AOM is used to shift the incident laser frequency far off-resonance. This ensures that no intracavity field builds up, effectively eliminating backscattering noise ($S_V^{\mathrm{bs}}$) and isolating the detector from any cavity induced dynamics. Next, the local oscillator (LO) driving signal to the EOM is disabled, thereby removing the RAM noise contribution ($S_V^{\mathrm{RAM}}$). Finally, to quantify the electronic noise floor ($S_V^{\mathrm{el}}$), we completely block the incident light while keeping the entire electronic readout chain fully operational. By subsequently unblocking the off-resonance light to measure the total voltage noise and subtracting the pre-calibrated dark noise ($S_V^{\mathrm{el}}$), we successfully extract the pure optical shot noise ($S_V^{\mathrm{sn}}$). This rigorous noise decomposition protocol allows us to predict the shot noise limit of the HUST-2 gyroscope.

By systematically varying the incident optical power $P_{\mathrm{in}}$ up to $1\,\mathrm{mW}$, we record 10 s continuous voltage output at a $20\,\mathrm{kHz}$ sampling rate for each power level. The corresponding root PSDs of these time-domain records, shown in Fig.~\ref{fig:shot_noise_results}b, exhibit flat white noise characteristics over a broad frequency range from 10 Hz to 10 kHz. Here, the measurement at $0\,\text{\textmu W}$ yields the pure electronic noise floor, $S_V^{\mathrm{el}}$. Subtracting this electronic noise floor from the total voltage noise at each incident power level isolates the pure optical shot noise. Furthermore, as depicted in the inset of Fig.~\ref{fig:shot_noise_results}b, we independently calibrate the frequency discrimination slope, $D$, which exhibits a linear dependence on $P_{\mathrm{in}}$. This power dependent slope serves as the essential conversion factor to translate the extracted voltage noise into its equivalent frequency noise.

Using these calibrated slopes and the Sagnac scale factor, $K$, we convert the isolated voltage noise into the equivalent rotational root PSD. Accounting for the uncorrelated shot noise contributions from both the CW and CCW locking loops, the measured fundamental shot noise limit of the HUST-2 gyroscope reaches an unprecedented $5.7(1) \times 10^{-13}\,\mathrm{rad/s/ \sqrt{Hz}}$ at an incident power of $1\,\mathrm{mW}$. As illustrated in Fig.~\ref{fig:shot_noise_results}c, this experimental result is in excellent agreement with the theoretical limit of $5.3 \times 10^{-13}\,\mathrm{(rad/s)/\sqrt{Hz}}$, calculated from Eq.~(\ref{eq:sn_realistic}) using parameters summarized in Methods, without any free parameters. We attribute the $\sim 7\%$ discrepancy to practical non-idealities in the PDH readout chain. Specifically, these include a suboptimal modulation depth, minor demodulation phase offsets, and residual electronic noise coupled into the shot noise limit.

\subsection{Noise budget of the HUST-2 PRG and state-of-the-art rotational resolution}

To approach the fundamental limit of this PRG, the technical noise components must be identified and suppressed. The continuously measured angular velocity $\Omega_{\mathrm{m}}(t)$ is a superposition of the true Earth rotation $\Omega_{\mathrm{E}}$ and several distinct error terms:
\begin{equation}
	\begin{aligned}
		\Omega_{\mathrm{m}}(t)
		&= \Omega_{\mathrm{E}}
		+ \delta\Omega_{\mathrm{env}}(t)
		+ \delta\Omega_{\mathrm{K}}(t) \\
		&\quad
		+ \delta\Omega_{\mathrm{meas}}(t)
		+ \delta\Omega_{\mathrm{freq}}(t),
	\end{aligned}
\end{equation}
where $\delta\Omega_{\mathrm{env}}(t)$, $\delta\Omega_{\mathrm{K}}(t)$, $\delta\Omega_{\mathrm{meas}}(t)$, and $\delta\Omega_{\mathrm{freq}}(t)$ represent environmental perturbations noise, scale factor fluctuations noise, measurement/readout noise, and residual noise from laser frequency locking, respectively.

HUST-2 is located in a deep underground laboratory in the Yujia Mountain, providing a stable thermal and vibration environment. The $32\,\mathrm{m}$ perimeter ring cavity operates in a vacuum of $10^{-5}\,\mathrm{Pa}$ to eliminate air refractive index fluctuations. Protected by two layers of passive thermal insulation, the daily temperature variation at the vacuum mirror chambers is within $0.5\,\mathrm{K}$. For vibration isolation, the four mirror chambers are mounted on independent isolated foundations, each supporting a 0.5 ton marble block to provide further passive damping. In addition to environmental control, geometric precision was ensured. A laser tracker was used during cavity assembly to accurately position the mirror chambers, limiting the maximum length deviation of the four 8 m long sides to less than $0.5\,\mathrm{mm}$. This tightly maintains the square geometry of the ring cavity and restricts the relative uncertainty of the Sagnac scale factor to a parts-per-billion level. Together, these environmental and structural measures effectively isolate the system from external perturbations, minimizing the environmental noise $\delta\Omega_{\mathrm{env}}(t)$ and suppressing the scale factor drift $\delta\Omega_{\mathrm{K}}(t)$.

Furthermore, for the frequency readout, an integrated monolithic beam-combining prism is employed to overlap the transmitted CW and CCW beams, as illustrated in Fig.~\ref{fig:sys}b. This monolithic design enhances the stability of the interference path and effectively suppresses measurement noise $\delta\Omega_{\mathrm{meas}}(t)$. The combined optical signal, reaching a power level of approximately $10\,\text{\textmu W}$, is directed to the $\mathrm{PD}_{\mathrm{Sagnac}}$ to generate a beat note waveform with a signal-to-noise ratio (SNR) exceeding $80\,\mathrm{dB}$. Following digitization at a sampling rate of $20\,\mathrm{kHz}$, the data is processed by a specialized frequency estimator that integrates single-tone estimation with the Hilbert transform. This approach precisely extracts the Sagnac beat frequency while resolving both high and low frequency signal dynamics.

Based on the noise decomposition in Eq.~(\ref{eq:somega_dis}), the frequency locking noise $\delta\Omega_{\mathrm{freq}}(t)$ in HUST-2 is mainly determined by the voltage noise terms in the PDH discrimination chain and their conversion through the discrimination slope $D$. A key design feature of HUST-2 is its large PDH discrimination slope, given by Eq.~(\ref{eq:pdh_slope}), which suppresses the conversion of voltage fluctuations into equivalent frequency noise. Benefiting from the $32\,\mathrm{m}$ perimeter and the high-$Q$ ring cavity, HUST-2 achieves an ultra narrow cavity linewidth of $\nu_{\rm c}=86.4(1)\,\mathrm{Hz}$. With an optimized injection power of $0.6\,\mathrm{mW}$, the resulting discrimination slope reaches $D\approx100\,\mathrm{mV/Hz}$, which effectively suppressing the equivalent noise induced by electronic noise $S_V^{\mathrm{el}}$. Additionally, wedged EOMs are used for passive RAM suppression, while low scattering coatings and the long ring cavity length reduce the backscattering coefficient to a parts-per-billion level, ensuring that $S_V^{\mathrm{bs}}$ remain negligible~\cite{zhong_backscattering_2025}.

\begin{figure}[htbp]
	\centering
	\includegraphics[width=\linewidth]{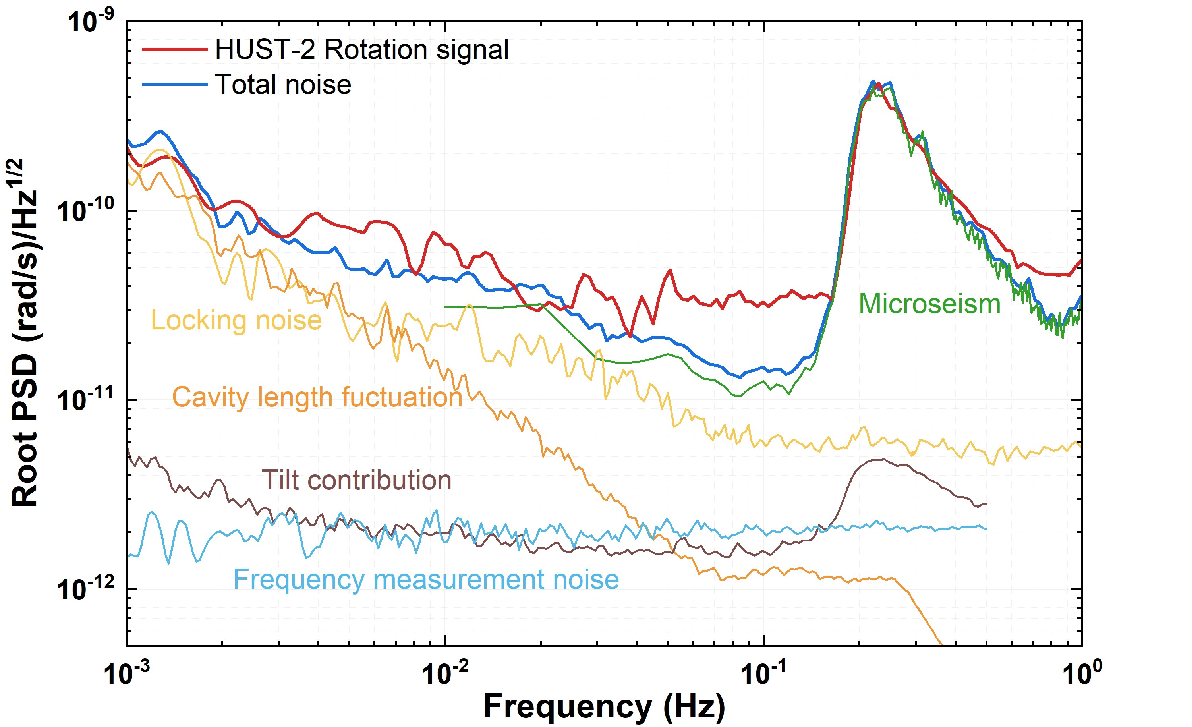}
	\caption{\textbf{Noise budget of the HUST-2 gyroscope.} The measured rotational resolution (red line) reaches $3 \times 10^{-11} \, \mathrm{(rad/s)/\sqrt{Hz}}$. The Total noise budget is blue line. The projected technical and environmental noise contributions are shown, including microseism (green), frequency locking noise (yellow), cavity length fluctuations (orange), ground tilt (brown), and frequency measurement noise (light blue).}
	\label{Fig3}
\end{figure}

\begin{figure*}[htbp]
	\centering
	\includegraphics[width=\linewidth]{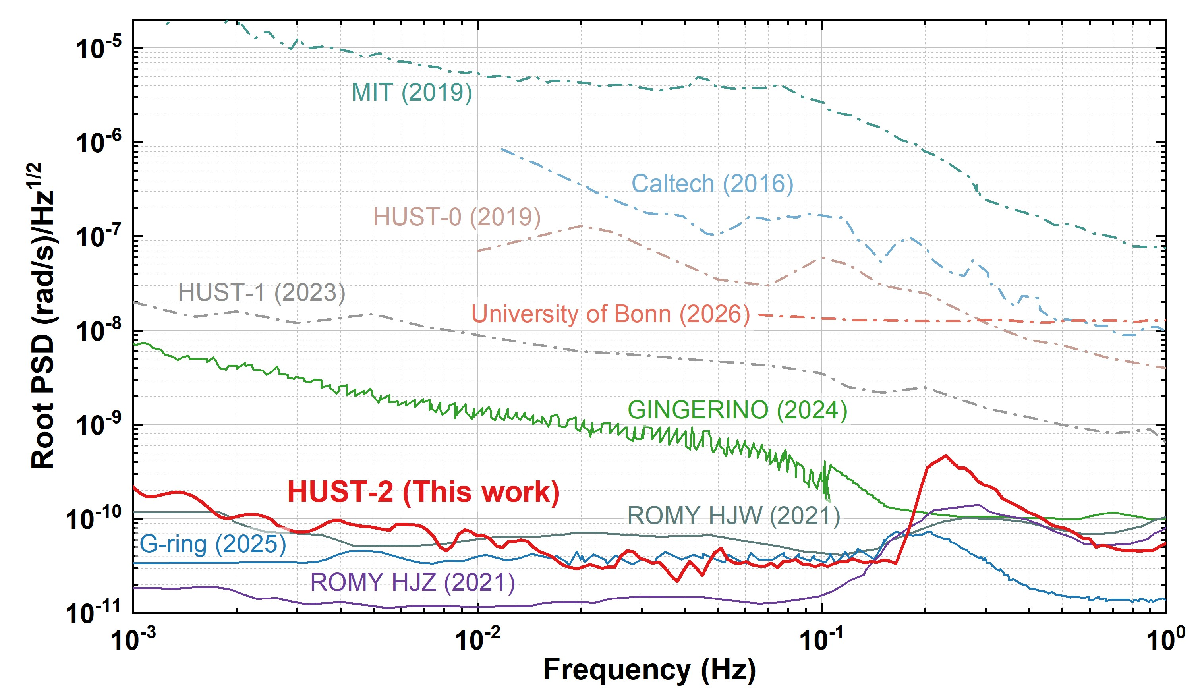}
	\caption{\textbf{Root PSD comparison of large-scale optical gyroscopes.} The performance of HUST-2 (red solid line) is benchmarked against the global state-of-the-arts. Dash-dotted lines represent existing PRGs, typically operating in the $10^{-7}$ to $10^{-9} \, \mathrm{(rad/s)/\sqrt{Hz}}$ range. Solid lines at the bottom denote premier RLGs. HUST-2 achieves a breakthrough rotational resolution of $3 \times 10^{-11} \, \mathrm{(rad/s)/\sqrt{Hz}}$, marking the first time a PRG has reached the performance echelon of the world's best RLGs.}
	\label{Fig4}
\end{figure*}

Following the systematic optimizations, HUST-2 achieves a rotational resolutions of $3 \times 10^{-11}\,\mathrm{(rad/s)/\sqrt{Hz}}$, shown as the red solid line in Fig.~\ref{Fig3}. To understand the system limits, the individual noise contributions are discussed in a top-to-bottom order. 
In the middle-frequency regime of $0.1$-$1\,\mathrm{Hz}$, the prominent peak observed around $0.2\,\mathrm{Hz}$ is not an instrumental noise limit, but a real geophysical signal corresponding to secondary oceanic microseisms~\cite{igel_romy_2021}. 
This is confirmed by the green line, which is obtained from a co-located seismometer and converted into an equivalent rotational signal using a proportionality coefficient $k \simeq 3\times10^{-4}\,(\mathrm{rad/s})/(\mathrm{m/s^2})$. 
The coefficient is determined from the amplitude ratio of the corresponding characteristic microseism peaks. In the low-frequency regime below $10^{-2}\,\mathrm{Hz}$, the frequency locking noise (yellow line) and the cavity length fluctuations (orange line) jointly dominate the long term drift. The locking noise, evaluated as the sum of two independent locking loops, exhibits a low-frequency drift. This drift is primarily attributed to the RAM noise, as only passive mitigation strategies are currently implemented. Similarly, while the cavity length shielded from previous passive isolation, the absence of active feedback stabilization results in fluctuations that currently limit the low-frequency performance. To evaluate scale factor related noise, an ultra-stable laser with a fractional frequency stability of $10^{-15}$ is employed to characterize the cavity length stability~\cite{zhang_long-term_2020}. By modulating the cavity length via a piezoelectric transducer (PZT) mounted on the mirror chamber and monitoring the induced Sagnac response, we determine a common-mode rejection ratio (CMRR) of $8 \times 10^8$, which quantifies the projection of cavity fluctuations onto the root PSD. At the lowest levels of the noise budget, the ground tilt contribution (brown line, derived from a co-located tiltmeter) and the frequency measurement noise (light blue line) remain completely negligible across the entire spectrum.

By combining the independently evaluated noise contributions, we obtain the total noise budget of the HUST-2 gyroscope, shown as the blue curve in Fig.~\ref{Fig3}. 
The summed noise reproduces the measured root PSD of the HUST-2 rotation signal over most of the investigated frequency band, indicating that the dominant noise sources have been properly identified and quantified. 
In particular, the calculated budget follows the measured spectrum in both the low-frequency region, where cavity length fluctuations and locking noise contributions are important, and the middle frequency region, where the microseismic peak dominates the spectrum. 
The good agreement between the blue curve and the measured root PSD confirms that the present rotational resolution of HUST-2 is mainly limited by the noise mechanisms included in the budget.

We benchmark the performance of HUST-2 against current state-of-the-art systems, as shown in Fig.~\ref{Fig4}. Historically, the rotational resolutions of PRGs, including our previous HUST-0/1 and other systems (dash-dotted lines), has typically been in the $10^{-7}$ to $10^{-9}\,\mathrm{(rad/s)/\sqrt{Hz}}$~\cite{martynov_passive_2019,korth_passive_2016,liu_large-scale_2019,feng_three-wave_2023,zenner_hansch-couillaud_2026}. By overcoming these long standing bottlenecks, HUST-2 represents a significant advancement for the PRG. With a measured rotational resolution of $3 \times 10^{-11}\,\mathrm{(rad/s)/\sqrt{Hz}}$ (red solid line), HUST-2 achieves, for the first time, a performance comparable to the best RLGs, such as ROMY, GINGERINO, and G-ring~\cite{igel_romy_2021,di_virgilio_noise_2024,schreiber_gyroscope_2025}.

Although the current performance is on par with leading RLGs, the much lower shot noise limit of the PRG indicates a clear potential for further improvement. As discussed above, the shot noise limit of HUST-2 reaches the $10^{-13}\,\mathrm{(rad/s)/\sqrt{Hz}}$ level, below the present rotational resolution of system. The reason why this limit is not yet reached is that laser gyroscope applications are mainly concerned with the low-frequency band below $1\,\mathrm{Hz}$, where technical and environmental noise sources are more complex and can readily exceed the shot noise limit. In HUST-2, the preceding noise analysis identifies the current limiting contributions as locking noise, cavity length fluctuations, and environmental perturbations. Further improvements will therefore require active RAM suppression, cavity length stabilization, advanced seismic isolation, and model subtraction of these contributions~\cite{descampeaux_new_2021,zhang_3_2020,brotzer_environment-related_2025}. Once these excess low-frequency noises are suppressed, the system is expected to approach its physical limit, providing a path toward next-generation rotational seismology and tests of fundamental physics.

\section{Discussion}\label{sec3}

We have demonstrated HUST-2, a $64\,\mathrm{m^2}$ giant PRG, with a rotational resolution of $3\times10^{-11}\,\mathrm{(rad/s)/\sqrt{Hz}}$. This represents an improvement of nearly two orders of magnitude over previously demonstrated PRGs, bringing PRG into the rotational resolution regime of state-of-the-art large-scale RLGs for the first time and closing the long-standing performance gap between the two architectures.

\begin{figure}[h]
	\centering
	\includegraphics[width=\linewidth]{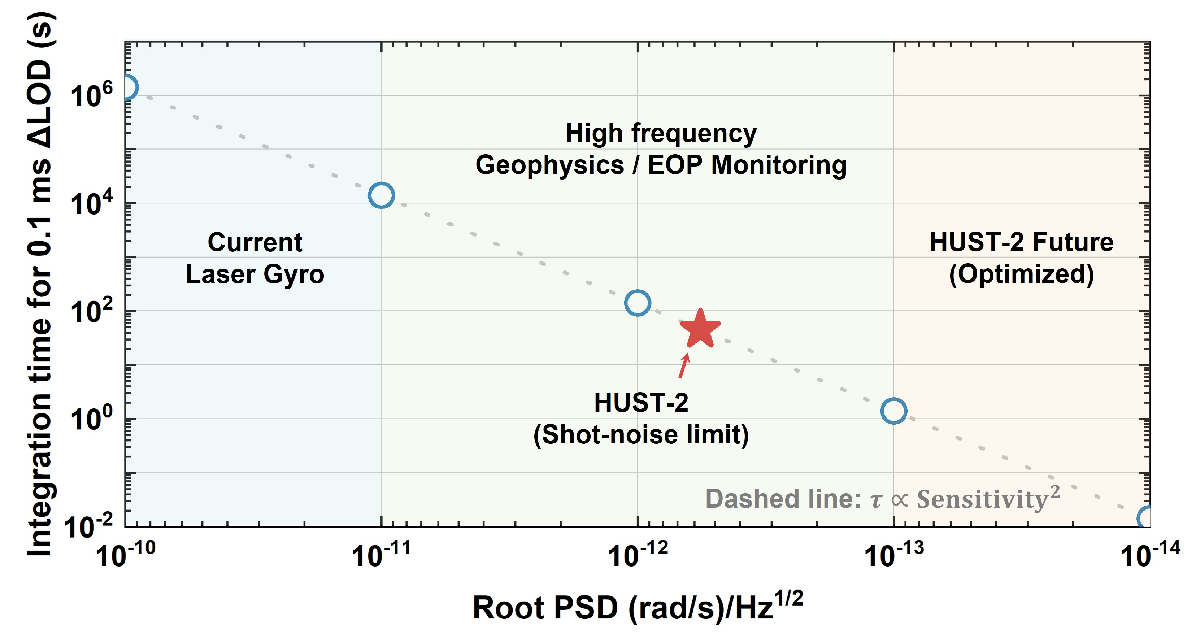}
	\caption{
		Required integration time for resolving a $0.1\,\mathrm{ms}$ $\Delta$LOD versus gyroscope rotational root PSD. The verified HUST-2 shot noise limit, $5.7(1)\times10^{-13}\,\mathrm{(rad/s)/\sqrt{Hz}}$, enables high-precision $\Delta$LOD measurements with a temporal resolution of about $10$-$100\,\mathrm{s}$.
	}
	\label{fig:LOD_resolution}
\end{figure}

Beyond the demonstrated system performance, we developed and experimentally validated a quantitative shot noise limit model for the PRG locking system. At $1\,\mathrm{mW}$ incident optical power, the measured shot noise limited reaches $5.7(1)\times10^{-13}\,\mathrm{(rad/s)/\sqrt{Hz}}$, in agreement with the model. This demonstrates that HUST-2 retains nearly two orders of magnitude of noise floor, with the present performance still limited by classical technical noise. As illustrated in Fig.~\ref{fig:LOD_resolution}, this shot noise limit enables the detection of $0.1\,\mathrm{ms}$ LOD variations with a temporal resolution of approximately $10$-$100\,\mathrm{s}$. The validated model further indicates that, by optimizing the photodetector saturation power, impedance matching, mode matching, and quantum efficiencies, large-scale PRGs can be pushed toward the $10^{-14}\,\mathrm{(rad/s)/\sqrt{Hz}}$ regime. This would extend to high frequency Earth rotation and EOP monitoring, and provide a route toward terrestrial tests of relativistic rotational effects.

Since a single-axis gyroscope measures only the local projection of the Earth rotation vector, full vectorial Earth-rotation metrology will require multiple gyroscopes with different orientations or locations. A second $10\,\mathrm{m}\times10\,\mathrm{m}$-scale PRG has therefore been constructed in the underground laboratory of the TianQin Research Center at the Zhuhai campus of Sun Yat-sen University and is currently reaching similar performance level. Together with HUST-2, it provides an experimental foundation for future large-scale gyroscope networks.

\section*{Methods}

\subsection*{Key Parameters of the HUST-2} 

The theoretical shot noise limit and the frequency-discrimination performance of the HUST-2 gyroscope depend on a specific set of geometric and optoelectronic parameters. These key fundamental and system parameters are summarized in Table~\ref{tab:parameters}.

\begin{table}[h]
	\centering
	\caption{Key technical parameters of the HUST-2 system.}
	\label{tab:parameters}
	\footnotesize
	\setlength{\tabcolsep}{3pt}
	\begin{tabular}{lcc}
		\hline
		\hline
		Parameter & Symbol & Value \\
		\hline
		Wavelength & $\lambda$ & $532\,\mathrm{nm}$ \\
		Cavity perimeter & $L$ & $32\,\mathrm{m}$ \\
		Enclosed area & $A$ & $64\,\mathrm{m^2}$ \\
		Cavity linewidth & $\nu_c$ & $86.4(1)\,\mathrm{Hz}$ \\
		Bessel product ($\beta \approx 1.08$) & $J_0 J_1$ & $0.343$ \\
		Mode-matching efficiency & $\alpha$ & $59.0(1)\%$ \\
		Impedance-matching coefficient & $\zeta$ & $0.72(1)$ \\
		On-resonance reflection ratio & $R$ & $85.8(1)\%$ \\
		On-resonance transmission ratio & $T$ & $5.3(1)\%$ \\
		Detector quantum efficiency & $\eta$ & $0.75$ \\
		Single-photon energy & $h\nu$ & $3.73 \times 10^{-19}\,\mathrm{J}$ \\
		\hline
		\hline
	\end{tabular}
\end{table}

To build the noise model, we experimentally characterized these system-specific parameters. First, by sweeping the laser frequency across the resonance and applying a Lorentzian fit to the transmitted power, we measured a narrow cavity linewidth ($\nu_c$) of $86.4(1)\,\mathrm{Hz}$, confirming the high quality factor of the ring cavity. When the laser is locked to the cavity, the on-resonance reflection and transmission ratios ($R$, $T$) were obtained by measuring the steady-state reflected and transmitted powers relative to the incident power. These ratios were then used to calculate the optical impedance-matching coefficient ($\zeta$). Furthermore, the spatial mode-matching efficiency ($\alpha$) was evaluated by measuring the fraction of the incident beam that successfully couples into the cavity's fundamental $\mathrm{TEM}_{00}$ mode. For the PDH locking setup, we used a Bessel function product ($J_0 J_1$) of $0.343$; this corresponds to the theoretical maximum of the frequency-discrimination slope, achieved by setting the modulation index close to the optimal value of $\beta \approx 1.08$. Finally, the photodetector quantum efficiency ($\eta$) was taken from the manufacturer's specifications.

\subsection*{Shot noise in transmission port}

In the PRG, the photon shot noise limited noise floor is determined by the optical detection processes. This noise arises at two distinct locations: the reflection port used for PDH frequency stabilization, denoted as $\mathrm{PD_{PDH}}$, and the transmission port used for Sagnac beat-note detection, denoted as $\mathrm{PD_{Sagnac}}$, as shown in Fig.~\ref{fig:sys}b. By explicitly modeling these noise contributions, we show that, for the present HUST-2 parameters, the overall shot noise limit is dominated by the frequency discrimination process within the PDH locking loop.

For direct heterodyne detection at the transmission port ($\mathrm{PD_{Sagnac}}$), the shot noise limited phase noise root PSD is given by:
\begin{equation}
	\sqrt{S_\phi^{\mathrm{sn}}}
	= \sqrt{\frac{2 h\nu}{\eta P_{\mathrm{det}}}},
	\label{eq:sphi_sn}
\end{equation}
where $P_{\mathrm{det}}$ is the optical power incident on the $\mathrm{PD_{Sagnac}}$. Converting this phase noise into frequency noise by multiplying by the Fourier frequency $f$, and dividing by the scale factor $K$, yields the rotation equivalent noise from the transmission port:
\begin{equation}
	\sqrt{S_\Omega^{\mathrm{sn,det}}}
	= \frac{\lambda L}{4 A}\,
	f \sqrt{\frac{2 h\nu}{\eta P_{\mathrm{det}}}}.
	\label{eq:somega_sn_det_general}
\end{equation}
Defining $P_{\mathrm{det}} = 2 T P_{\mathrm{in}}$, where $T$ is the on-resonance transmission ratio, and accounting for the transmission through the two loops, Eq.~(\ref{eq:somega_sn_det_general}) becomes
\begin{equation}
	\sqrt{S_\Omega^{\mathrm{sn,det}}}
	= \frac{\lambda L}{4 A}
	\,f \sqrt{\frac{ h\nu}{\eta T P_{\mathrm{in}}}}.
	\label{eq:somega_sn_det}
\end{equation}

Their ratio is computed to evaluate the impact of this beat-note shot noise relative to the PDH frequency discrimination noise. Using the parameters of the HUST-2 gyroscope listed in Table~\ref{tab:parameters}, we obtain:
\begin{equation}
	\begin{aligned}
		\frac{\sqrt{S_\Omega^{\mathrm{sn,det}}}}
		{\sqrt{S_\Omega^{\mathrm{sn}}}}
		&=
		\frac{4J_0 J_1 \alpha (1-\zeta)f}{\nu_c\sqrt{RT}} \\
		&\approx 1.2\times 10^{-2} \left(\frac{f}{1\,\mathrm{Hz}}\right).
	\end{aligned}
	\label{eq:ratio_det_ideal}
\end{equation}
This analysis shows that, within the sub-hertz observation band relevant for Earth rotation sensing, the shot noise associated with beat-note detection is more than two orders of magnitude lower than the locking limited noise. Consequently, the beat-note shot noise is negligible in the present noise budget of the HUST-2 PRG.

\section*{Acknowledgements}

The authors are grateful to all colleagues and collaborators for helpful discussions and technical assistance.

\section*{Funding}

This work was supported by the National Natural Science Foundation of China 
(12574530), and the National Key Research and Development Program of China 
(2022YFC2204002 and 2022YFB3904001).

\bibliographystyle{nsr}
\bibliography{nsr_sample}

\end{document}